# Performance of OFDM System against Different Cyclic Prefix Lengths on Multipath Fading Channels

Muhammad Firdaus[1], and Yoedy Moegiharto[1]
[1]Department of Telecommunication Engineering,
Electronic Engineering Polytechnic Institute of Surabaya, Indonesia

muhammadfirdaus.mufi@gmail.com, ymoegiharto@pens.ac.id

*Abstract*—Orthogonal Frequency Division Multiplexing (OFDM) is a transmission technique that uses several subcarrier frequencies (multicarrier) that are perpendicular to each other (orthogonal). This OFDM modulation technique effectively eliminates Intersymbol Interference (ISI) on a channel caused by the effects of multipath fading. To overcome this weakness, OFDM uses a guard interval (cyclic prefix) inserted in its transmission. This paper analyses the effect of different cyclic prefix lengths on OFDM performance on multipath channels using Matlab. Here, the cyclic prefix length as the main parameter is varied for the different number of subscribers. Meanwhile, Bit Error Rate (BER) and Signal Noise Ratio (SNR) values obtained at the receiver are used to see the system's performance. Based on our simulation results, the best value of BER is obtained using a cyclic prefix length of 1/4 with a Fast Fourier Transform (FFT) size of 512.

## I. Introduction

In line with the rapid development of information technology and telecommunications, digital wireless communication systems are required to provide high-speed data services (high data rate) with reliable QoS. One technique that can be relied upon to provide high-speed data services is the Orthogonal Frequency Division Multiplexing (OFDM) multicarrier modulation technique. OFDM has long been used as a solution to serve as the primary interface on several wireless systems such as W-LAN (IEE 802.11), Digital Audio Broadcasting, Digital Video Broadcasting and Wi-Max systems (IEE 802.16).

In addition, OFDM multicarrier modulation technique can be used to overcome the effects of multipath fading that can cause Intersymbol Interference (ISI). In order to eliminate the negative effect of ISI, a cyclic prefix (CP) is used. However, the use of CP makes the symbols sent longer, resulting in a decrease in the symbol rate and channel capacity. To overcome the decrease in channel capacity, send symbols in OFDM using multiple carriers and frequencies. However, the use of multiple carriers and multiple frequencies creates intercarrier interference (ICI). So to avoid and reduce ICI, the carriers are made orthogonal to each other.

In this paper, we study and simulate the effect of different cyclic prefix lengths for different subscriber numbers on OFDM system performance by looking at the Bit Error Rate (BER) value in the Matlab simulator. Our main contribution can be summarized as follows.
1. We implemented OFDM system performance on Matlab simulator
2. We analyze the performance of the OFDM system by using different cyclic prefix lengths for different numbers of subscribers
3. We analyze OFDM performance parameters, namely BER and Eb/No, obtained from simulation results.

The rest of this paper is arranged as follows: we briefly explain the background knowledge and concept of OFDM, multipath fading, ISI, and cyclic prefix in Section 2. Section 3 explains the OFDM system's system design, data retrieval design, and data processing processes. Then, we examine the results and considerations in Section 4. Finally, Section 5 concludes this paper.

## II. Background

Wireless channels are a fundamental part of understanding the operation, design and analysis of any wireless system as a whole, such as in cellular communication systems, radio paging or mobile satellite systems. Moreover, the wireless channel is the main factor limiting the performance of wireless communication systems. The transmission distance between transmitter and receiver can vary from Line of Sight (LOS) to covered by obstructions such as buildings, hills and trees. Unlike wired channels which are fixed and predictable, radio channels are random and cannot be analyzed easily. Further, the speed with which the user moves affects how quickly the signal level fades.

The characteristics of wireless channels can be grouped into two major groups: large-scale fading and small-scale fading. Fading itself means the phenomenon of signal power fluctuations (tends to weaken) received due to the propagation process of radio waves. The received signal is random in time and space [1]. Large scale fading is caused by the phenomenon of free space propagation, namely direct path, diffraction and scattering, while small scale fading is caused by reflection by objects on the wireless channel or is called a multipath phenomenon [3]. The phenomenon of small-scale fading is used to describe fluctuations in the amplitude of a radio signal over a speedy period and a short distance. Fluctuations occur not because of the path loss effect, which is a distance function. Hence, it is concluded through signal fluctuations even if the user is silent (there is no change in distance). Fading is generated by the interference of two or more multipath signals at the receiver, which causes a phenomenon of fluctuation in the received signal power (which tends to weaken) due to the propagation process of radio waves. The received signal is random in time and space.

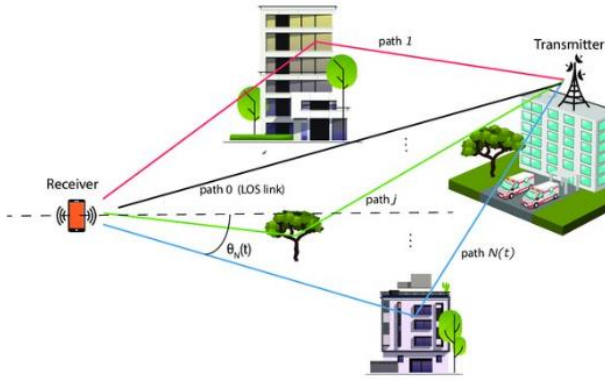

Fig. 1. Multipath propagation [9].

*A. Multipath fading channel*

The effect of signal reflections causes multipath fading phenomena. Moreover, small-scale fading describes signal power fluctuations that change rapidly in a short period or over a short distance—generally caused by interference between two or more replicas of a signal emitted from a transmitter and arrives at one receiving point at different times and directions of arrival. The replicas of these signals are called multipath signals, at the receiver will be added up and produce a signal with an enlarged phase and amplitude (widened signal shape). The following factors affect small-scale fading[2]:

- *Multipath propagation:* As shown in Figure 1, the reflecting objects in the channel produce a signal energy dissipation phenomenon. The energy dissipation of the signal is in the form of amplitude, phase, and time dissipation. So that the received signal will have various versions, namely a direct version and a delayed version with random amplitude and phase variations for each multipath component. In most communication systems, this can result in ISI (Intersymbol Interference).
- *Receiver speed:* The relative movement between the base station and the receiver will result in a random modulation frequency due to the difference in Doppler Shift on each multipath component. Doppler Shift will be positive or negative depending on whether the receiver is moving toward or away from the base station.
- *Speed of surrounding objects:* If the object on the radio channel moves, there will be a Doppler Shift change with time for each multipath signal. The effect of the object's movement will be dominant when the object moves faster than the receiver and vice versa.
- *Bandwidth signal:* For a transmission bandwidth greater than the coherent bandwidth of the signal, the signal will experience distortion. This signal distortion is the result of frequency selective fading, a unique characteristic of the transmission channel itself.

*B. Orthogonal Frequency Division Multiplexing (OFDM)*

Orthogonal Frequency Division Multiplexing (OFDM) is a transmission technique that uses several frequencies (multicarrier) that are perpendicular to each other (orthogonal). Each subcarrier is modulated by conventional modulation technique at a low symbol ratio. OFDM has several advantages when compared to multicarrier FDM techniques, one of which is more efficient bandwidth [2]. The process is the same, but the difference is the use of orthogonal subcarriers in each subchannel. In principle OFDM divides high-speed serial data into several low-speed parallel data, which then each parallel data is modulated by orthogonal subcarriers. This OFDM subcarrier orthogonality causes the spectrum between subcarriers to be allowed to overlap so that bandwidth usage will be more efficient [4]. Figure 2 illustrates the difference between the non-overlap (conventional) multicarrier technique and the orthogonal multicarrier modulation (OFDM) technique. The OFDM multicarrier technique can save bandwidth by almost 50%. Orthogonality is obtained by setting the distance between the carriers to 1/T, where T is the symbol period.

*C. Intersymbol Interference (ISI)*

Inter-symbol Interference (ISI) is an unavoidable problem in wireless communication systems. Each transmitted signal will experience multipath so that the receiver will receive a signal which incidentally results from the accumulation of the same message, which is delayed and with varying power. Intersymbol Interference (ISI) is a form of signal distortion where one symbol interferes with another distortion. Hence, it can occur due to signal reflection, which causes the reception of the information signal to repeat at different times or experience delays [5]. The presence of ISI will cause an error in receiving bits of information on the receiving side. One of the causes of Intersymbol Interference (ISI) interference is multipath propagation, where the transmitter signal reaches the receiver via many different paths [8]. Thus, it means that some or all of a particular symbol will propagate to the next symbol, thereby interfering with the correct detection of symbols. The solution to deal with ISI is to add an equalizer on the receiving end.

*D. Cyclic Prefix*

Cyclic Prefix (CP) is a mechanism for adding symbols by taking several symbols at the end of the IFFT frame to be inserted at the beginning of the frame, which results in a guard period. The Cyclic Prefix comes before the placement of the input symbol. Hence, it is so that when signals from several subcarriers equipped with cyclic prefixes are sent to a channel that is frequency selective fading, the signal sent can avoid Intersymbol Interference (ISI) and can be recovered properly by the receiver even though there is still a fading effect channel to affect the signal sent [6].

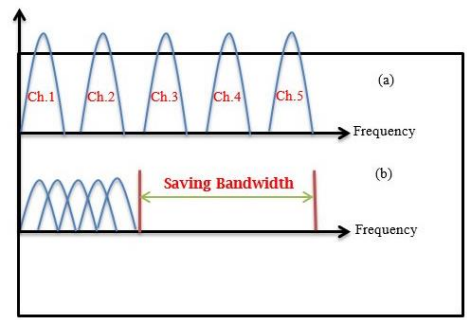

Fig. 2. Multipath propagation [10].

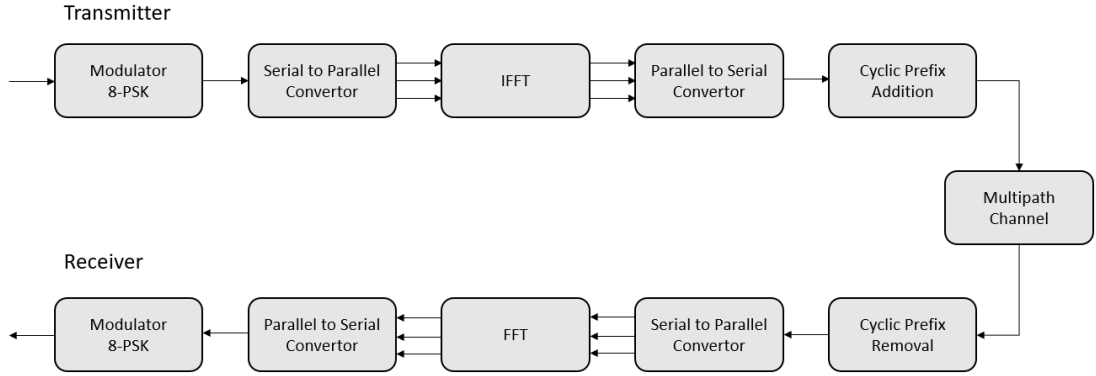

Fig. 3. Block diagram of OFDM system.

The main purpose of adding Cyclic Prefix to the data is to maintain or protect the orthogonality of the input symbol within a certain period from the phenomenon [7]. Because if there is noise, it is not the data symbol that has an error but the Cyclic Prefix bit. ISI can be omitted if the cyclic prefix length exceeds the channel's maximum excess delay.

III. OFDM SYSTEM DESIGN AND IMPLEMENTATION

This section discusses the design of the system, the manufacturing procedure, and the results displayed from the system. Making the system starts from data generation, modulation process, the data conversion process from serial to parallel, IFFT process, addition of a cyclic prefix, addition of multipath fading channel, deletion of cyclic prefix, the data conversion process from parallel to serial, FFT process, demodulation process and finally the output in the form of BER graph. After the system is made as a whole, testing and data collection are carried out to be used as material for analysis. In constructing an OFDM system, several parameters are needed to run the system. The parameters used are as follows:

- Number of FFTs used 64, 128 and 256
- Length of cyclic prefix 1/16, 1/8, 1/4 and 1/2
- The modulation used is MPSK (8-PSK)
- The channel used is a multipath channel
- The number of symbols used is 1000 bits

This paper discusses the effect of cyclic prefix length on OFDM performance on multipath channels with MPSK (8-PSK) modulation. Hence, to show the system performance, it is indicated by the error probability of Bit Error Rate (BER) as a function of SNR. Figure 3 shows the block diagram of the system design. The simulation stages of the OFDM system will be divided into three, namely: Transmitter (Tx), Channel (multipath), and Receiver (Rx). The detailed implementation of each block is shown in Figure 4 and described as follows.

- **Random data generation:** Input data is generated from bits of information that will be sent randomly. The generation of these bits is generated randomly by Matlab by using the randi command in Matlab. Hence, the generated bits are binary data bits 0 and 1. A series of data bits can be seen as shown in Figure 4.a. The input data is obtained from the multiplication of the number of subcarriers, the number of symbols, and the marry modulation used. Here, the number of bits used is 1000.

- **Modulator 8-PSK (Phase Shift Keying):** The modulation process is setting the parameters of the carrier signal with a high frequency according to the information with a lower frequency. At this stage, the input data bits that have been generated are then modulated using a PSK modulator. PSK modulation used is 8-PSK. For 8-PSK modulation, the number of n used is n = 3 to produce a phase difference of eight or M = 8. The 8-PSK modulation has eight phase difference positions of 45° each with 3 bits per symbol. The constellation results from 8-PSK modulation can be seen in Figure 4.b.

- **Serial to Parallel (S/P) converter:** This serial to parallel input is in the form of complex number data. The result of serial to parallel conversion is in the form of a matrix of bits with the number of rows stating the number of subcarriers used and the number of columns stating the number of data symbols sent to each subcarrier.

- **IFFT (Invers Fast Fourier Transform):** The next process is that the signal is applied to the IFFT for making OFDM symbols. In this IFFT, subcarrier frequencies are generated, which are orthogonal. The function of IFFT is to convert from the frequency domain to the time domain. The IFFT input is in the form of complex number data. The data will be sent for every N symbol, where N is the number of subcarriers used. The number of subcarriers used in this simulation is 64, 128, and 256 subcarriers. If the result of the modulation process is in the form of serial data, then the input for the IFFT process is in the form of parallel data.

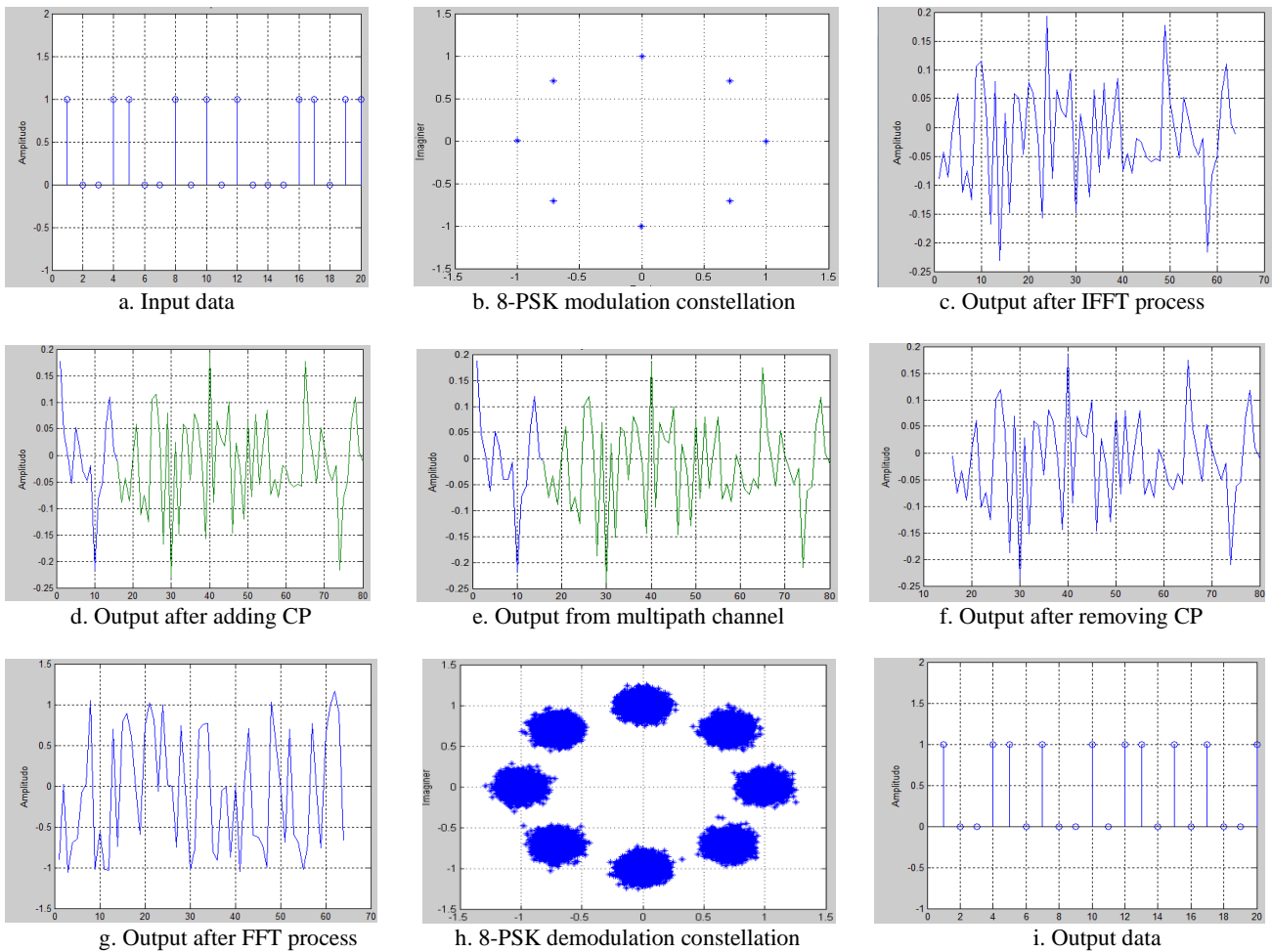

Fig. 4. Block diagram of OFDM system.

Therefore before the IFFT process, the data is changed from serial to parallel. On the other hand, the use of IFFT is also intended to replace the function of the analog modulator in OFDM, making the system difficult to realize because the hardware will be very complicated. Consequently, IFFT is used to replace the analog modulator function in OFDM. In figure 4.c. the signal is shown after the IFFT process using the number of 64 subcarriers.

• **Cyclic prefix addition:** In OFDM, multipath delay can be overcome by using a guard interval or a cyclic prefix. The condition to avoid ISI is that the duration of the cyclic prefix must be longer than the duration of the multipath delay. The guard interval consists of a copy of the end of the OFDM symbol because the receiver will later integrate each multipath through an integer number of sinusoidal cycles when demodulating OFDM with FFT. In this simulation, CPs with varying lengths are used to determine the effect of the cyclic prefix length, namely 1/2, 1/4, 1/8, and 1/16 of the OFDM symbol before the cyclic prefix is given. In addition, a cyclic prefix is inserted to eliminate the effect of ISI (intersymbol interference). Figure 4.d. shows a signal of one OFDM symbol using CP 1/4. The blue part of the signal represents the cyclic prefix which is a copy of the last 25% OFDM symbol, and the green part is the OFDM symbol.

• **Parallel to Serial (P/S) converter:** After adding the Cyclic Prefix, the input data bits are converted from parallel data input to serial. Parallel input data, consisting of only one row and several columns, is then converted into serial data, namely data in many rows and columns. The change in the input form is because serial data represents an analog signal. In this sense, the data bits can be transmitted if it is in serial form.

• **Multipath channel:** Data that has been processed at the transmitter will then be sent via a multipath fading channel. Therefore, the received signal results from a superposition of reflected signals with different amplitudes and phases. In the multipath fading channel, it is obtained through fluctuations in the amplitude of the signal rapidly in a certain period, which is caused by the receipt of two or more same signals by the receiver due to a large number of signal paths. So, noise and multipath fading are added to the data sent in this process. Multipath fading is added by multiplying the data by fading, then adding noise. Channels are randomly generated at different times - meaning each transmitted symbol will be multiplied by a randomly varying complex number. Figure 4.e. shows the OFDM signal after passing through the channel by simulating the AWGN channel using 64 subcarriers.

- **Cyclic prefix removal:** On the receiver side, the data bits in serial data are converted into parallel data. Then, the Cyclic Prefix bit will be omitted so that the original input bit can be processed. The information bits still accompanied by the Cyclic Prefix bit are separated by the Cyclic Prefix bit using CP Remover. The way the Cyclic Prefix (CP) Remover works is the opposite of the Cyclic Prefix on the transmitter block. If CP data is added to the transmitter block, then in the CP Remover process at the receiver, the data accompanied by CP in the previous process is removed so that the original input data symbol is a time-domain analog signal. Figure 4.f. shows a signal after cyclic prefix removal.

- **FFT (Fast Fourier Transform):** The Fast Fourier Transform (FFT) process is the opposite of the IFFT process at the transmitter. FFT has a function to parse OFDM symbols. FFT changes from the time domain to the frequency domain for the later 8-PSK demodulation process before finally getting the output data in the form of binary information data. The output of the FFT is shown in Figure 4.g.

- **Demodulator 8-PSK:** Demodulation is done to get back the information signal which is superimposed on the carrier signal. This process means the return of the information signal from the signal that the 8-PSK modulator has modulated on the transmitter. Figure 4.h. shows the output signal and constellation of the 8-PSK demodulator. After performing the 8-PSK demodulation process, binary data, which is the initial information signal sent, will be obtained. The result can be seen in Figure 4.i.

## IV. PERFORMANCE MEASUREMENT AND NUMERICAL RESULTS

In this chapter, an analysis of the system simulation results is carried out. The results are represented in the form of a Bit Error Rate (BER) graph to determine the system's performance. BER is found by dividing the number of incorrect bits by the number of bits sent. The analysis is carried out based on the FFT size parameter, the cyclic prefix length parameter, and the number of subcarrier parameters, reffered to Table 1. Meanwhile, the

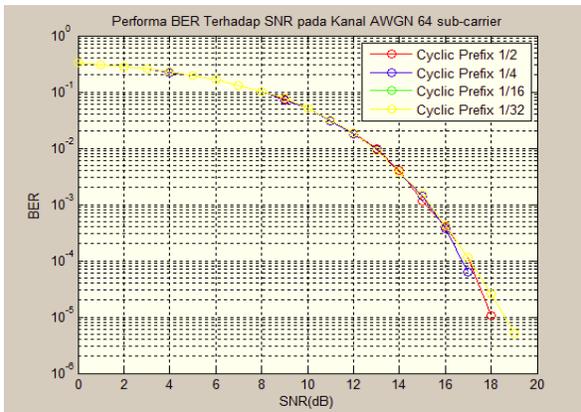

a. Results of comparing BER to SNR with a size of 64 FFT

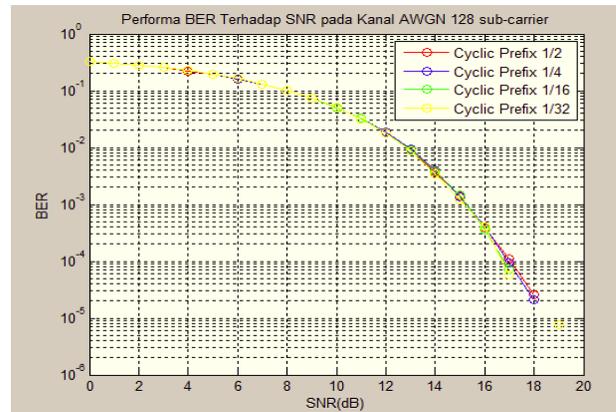

b. Results of comparing BER to SNR with a size of 128 FFT

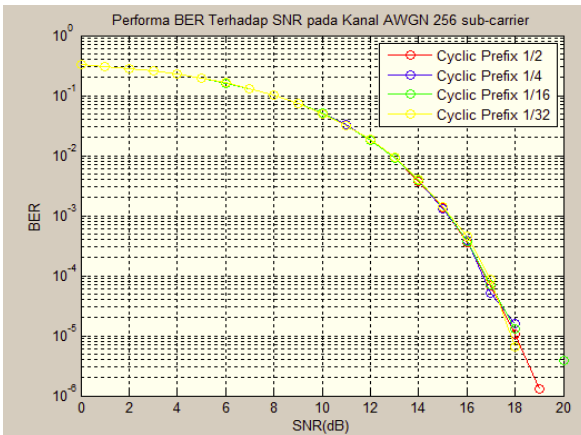

c. Results of comparing BER to SNR with a size of 256 FFT

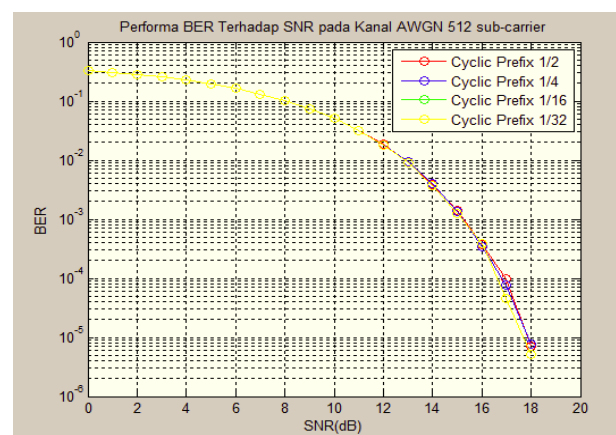

d. Results of comparing BER to SNR with a size of 512 FFT

Fig. 5. The effect of cyclic prefix length (CP) on the OFDM system with size variations.

results of the effect of cyclic prefix length (CP) on the OFDM system with size variations is shown in Figure 5.

Table 1. Simulation parameters.

| No | *Cyclic prefix* | Besar FFT |
|----|-----------------|-----------|
| 1  | 1/2             | 64        |
| 2  | 1/4             | 128       |
| 3  | 1/16            | 256       |
| 4  | 1/32            | 512       |

- **CP length with 64 FFT size**

In Figure 5.a, the simulation results show the effect of cyclic prefix length on OFDM system performance with an FFT size of 64. In this simulation, the AWGN channel is used. The simulation results show 4 curves, each of which shows the performance of using varying cyclic prefix lengths. The length of the cyclic prefix used is 1/2, 1/4, 1/16, and 1/32 of the length of the OFDM symbol used. Moreover, it shows that using the 1/32 cyclic prefix produces the worst beer error rate. The simulation using the 1/32 cyclic prefix has resulted in a BER below $10^{-3}$. If the standard BER for permitted voice communication is $10^{-3}$, then the 1/32 cyclic prefix can be used. Likewise, the cyclic prefix lengths of 1/16.1/4 and 1/2 produce a BER that exceeds $10^{-3}$. For CP 1/16 length, the resulting BER is better than CP 1/32 length. While the length of CP 1/4 produces the best BER among the four other cyclic prefix lengths, there is no significant difference numerically.

- **CP length with 128 FFT size**

Figure 5.b shows the simulation results of the effect of cyclic prefix length on OFDM system performance with an FFT size of 128. In this simulation, the AWGN channel is used. The simulation results show 4 curves, each of which shows the performance of using varying cyclic prefix lengths. The length of the cyclic prefix used is 1/2, 1/4, 1/16, and 1/32 of the length of the OFDM symbol used. Moreover, it shows that using the cyclic prefix 1/16 produces the worst beer error rate. Even so, the simulation using the cyclic prefix 1/16 has resulted in a BER below $10^{-3}$, so it can be used for voice communication. Of the four cyclic prefix lengths used in this system with an FFT size of 128, they all produce almost the same BER values and do not differ much. The lengths of CP 1/2 and 1/4 are still better than those of CP 1/16 and 1/32 when viewed from the resulting BER.

- **CP length with 256 FFT size**

Figure 5.c. shows the simulation results of the effect of cyclic prefix length on OFDM system performance with an FFT size of 256. In this simulation, the AWGN channel is used. The simulation results show 4 curves, each of which shows the performance of using varying cyclic prefix lengths. The length of the cyclic prefix used is 1/2, 1/4, 1/16, and 1/32 of the length of the OFDM symbol used. Moreover, it can be seen that the use of the cyclic prefix 1/16 produces the worst beer error rate. Even so, the simulation using the cyclic prefix 1/16 has resulted in a BER below 10-3, so it can be used for voice communication. Of the four cyclic prefix lengths used in this system with a size of 256 FFT, they all produce almost the same BER values and do not differ much. The lengths of CP 1/2 and 1/4 are still better than those of CP 1/16 and 1/32 when viewed from the resulting BER.

The use of a cyclic prefix in the OFDM system with an FFT size of 256 produces a better bit error rate value when compared to an OFDM system that uses an FFT size of 64 and 128. Hence, it is because the size of the FFT will affect the length of the symbol formed by the OFDM system. The duration of the OFDM symbol for FFT 256 size is longer than the symbol duration for OFDM with sizes 64 and 128. So that when the symbol is sent, the effect of multipath channels can be minimized because the allocation of data bit placement is longer.

- **CP length with 512 FFT size**

Figure 4.4 shows the simulation results of the effect of cyclic prefix length on OFDM system performance with an FFT size of 512. In this simulation, the AWGN channel is used. The simulation results show 4 curves, each of which shows the performance of using varying cyclic prefix lengths. The length of the cyclic prefix used is 1/2, 1/4, 1/16, and 1/32 of the length of the OFDM symbol used. Moreover, it shows that the four types of curves representing cyclic prefixes with lengths of 1/2,1/4.1/16, and 1/32 produce a BER value that is almost the same at SNR 17dB to SNR 20dB. Compared with the use of FFT sizes of 64, 128, and 256, FFT 512 improves the performance of the cyclic prefix 1/32 and 1/16. Even the BER values in some SNRs are almost the same as the length of the cyclic prefixes 1/4 and 1/2.

In the OFDM system that uses FFT 512, it produces a better BER value when compared to the use of FFT 64,128, and 256. Hence, it is because using FFT 512 can minimize the effects of multipath channels. The subcarrier produced by FFT 512 is more than twice that of FFT 256, four times that of FFT 128, and 8 times that of FFT 64. That way, the allocation of data bits sent is much more because the duration of the OFDM symbol for FFT 512 is longer than that of FFT 512. the other size. With the same number of input data bits, there are several empty slots of the subcarriers provided by FFT 512. So that the multipath channel effect does not affect the data bits sent but rather the remaining slots of the provided subcarriers, it can be said that the use of a larger FFT size will result in a better bit error rate for OFDM system performance.

V. CONCLUSION

We have presented the effect of different cyclic prefix lengths on OFDM performance on multipath channels using Matlab. Here, we used AWGN channel with 8-PSK modulation. Based on simulation results, we can found that the use of cyclic prefix can improve the performance of OFDM system. The cyclic prefix length used should be more than 1/4. The larger the cyclic prefix, the better the Bit Error Rate (BER) value. Vice versa, the smaller the value of the cyclic prefix, the BER value will be greater or worse. On the other hand, using a larger FFT (Fast Fourier Transform) size will increase OFDM performance. Because the greater the value of FFT, the duration of the OFDM symbol will be longer, so the allocation of data bits sent will be more. In this final project simulation, the best BER value is shown using an FFT of 512.